\documentclass[11pt]{article}
\usepackage{amsmath,amssymb}
\usepackage[dvips]{graphicx}

\begin{document}

\begin{titlepage}

\begin{center}
\hfill OU-HET 472 \\
\hfill UOSTP-04102 \\
\hfill hep-th/0404104
\vspace{2cm}

{\Large\bf Phase Moduli Space of Supertubes}
\vspace{1.5cm}

{\large
Dongsu Bak,$^a$ Yoshifumi Hyakutake$^b$ and  Nobuyoshi Ohta$^b$}

\vspace{.7cm}

$^a${\it
Physics Department, University of Seoul, Seoul 130-743, Korea
\\ [.4cm]}
$^b${\it  Department of Physics, Osaka University,
Toyonaka, Osaka 560-0043, Japan\\
[.4cm]}
({\tt dsbak@mach.uos.ac.kr, hyaku@het.phys.sci.osaka-u.ac.jp,
ohta@phys.sci.osaka-u.ac.jp})

\end{center}
\vspace{1.5cm}

\begin{quote}

We study possible deformations of BPS supertubes keeping their
conserved charges fixed. We show that there is no flat direction to
closed supertubes of circular cross section with uniform electric
and magnetic fields, and also to open planar supertubes.
We also find that there are continuously infinite flat deformations to
supertubes of general shape under certain conditions.

\end{quote}

\end{titlepage}
\setcounter{page}{2}

\section{Introduction}

Supertubes are the bound states of D0-brane, fundamental strings and
D2-branes preserving  1/4 supersymmetries in the type IIA superstring
theory~\cite{mateos,klee}. The original solutions were found to have
a circular cross sectional shape. It was realized that the
1/4 BPS elliptic supertubes~\cite{karch} are possible and the cross
sectional shape can be in fact completely arbitrary~\cite{ng}.
Since then there have appeared many literatures
discussing properties of the supertubes~\cite{cho}--\cite{lunin}.

The arbitrariness of the cross sectional shape in the target space
is unusual from the view point of the conventional D-branes.
The world-volume magnetic field representing D0-brane density may also
be arbitrary along the cross sectional contour while the density
of F1 is completely determined  by the magnetic field and the shape.

The dynamics behind the arbitrariness does not seem to be quite understood.
Note that the energy of supertubes as a 1/4 BPS configuration is solely
determined by the total numbers of D0 and F1 strings.
One may then ask if the change  in the cross sectional area is related
by a flat direction or not. This is a simple minded question. Or can they be
fluctuating from one shape to another freely? To answer this type of
questions, figuring out the landscape of energy of supertubes
would  be enough. In the description of supertubes using the
Dirac-Born-Infeld action, the supertubes have a nonzero kinetic contribution.
Namely the world-volume electric field and the momentum density are
nonvanishing for any 1/4 BPS supertubes. Thus we are interested in
the phase space landscape of energy. The landscape is in general complicated.
The number of D0-branes and fundamental strings of the system are
conserved and cannot be changed by the dynamical processes.
With fixed D0 and F1 numbers, we shall show that any area change costs
energy for uniform circular supertubes.
When the cross sectional area becomes larger or smaller, the energy
will be shown to be increasing always.

There are further conserved quantities of the system. The total angular
momentum and total linear momentum are conserved of course.
Having statistical applications in mind, we are
particularly interested in the phase space landscape of
energy with all the conserved quantities fixed.

For fixed charges of D0, F1 and the angular momentum, there is
a minimum energy region of the phase space. We shall show that
all the points in the region are connected dynamically by flat
directions and 1/4 of supersymmetries are preserved throughout the region.
This definition of the region  is analogous to the usual moduli space
but we are working in the phase space instead of the configuration
space. One may call this as phase moduli space; its volume corresponds
to semi-classically to the ground state degeneracy with the fixed  charges.
In this note, we shall explore the detailed structure of the phase
moduli space of the supertubes. 
We consider both the cases of closed and open~\cite{karch} supertubes.
For the closed case, we shall show that
there is no flat direction if the circular supertubes have uniform
distribution of D0 and F1 distributions. This corresponds to the
case where $Q_0 Q_1=J$ where $Q_0$ is the number of D0 in the
unit length along the direction of axis, $Q_1$ is the
total number of fundamental strings and $J$ corresponds to the angular
momentum. Hence the region is described by a single point
having zero volume. On the other hand, if $Q_0 Q_1 > J$, the change
in the shape of the tubes and the distribution of D0-branes are flat
directions. We shall give the details of all possible flat directions.

For the open supertubes, we shall show that the planar D2
with uniform D0 and F1 distribution again does not allow any flat
directions.  Though we shall prove the above claim
explicitly, it may be understood as follows. Let us consider the circular
supertube with  large radius with a uniform world-volume
density of D0 branes. In the limit of infinitely large
radius, the problem becomes essentially that of open supertube
if we focus on any large but finite part of the closed supertube.
Thus it is clear that the open  planar supertube
with uniform D0 and F1 densities does not allow any flat directions
except the trivial overall translation in the transverse space.
The case of arbitrary open supertube may be understood
similarly from the closed supertubes in the large size limit.

This paper is organized as follows. In Section 2, we begin with
DBI action of supertubes and discuss the BPS equations of
supertubes with arbitrary cross sectional shape. We shall
then discuss the simple example of radius changing landscape
of circular supertubes with fixed numbers of D0 and F1.
In Section 3, we shall show that circular supertube with uniform
D0 and F1 densities does not allow any flat directions except
the trivial overall translation of the tubes.
Section 4 deals with all possible flat directions by the
explicit construction for the cases of $Q_0 Q_1 > J$.
This clarifies the detailed structures of the phase moduli  space of
supertubes. Last section is devoted to the discussions.

\section{BPS Equations and Conserved Charges for a General Supertube}

We start with the discussion of BPS equations and conserved charges for
a general supertube. The result will be used to examine the flat
directions under the condition that the charges are conserved.

The supertube is a bound state of D2-D0-F1 in the flat spacetime.
We parametrize the flat target spacetime as
\begin{alignat}{3}
ds_{10}^2 &= -dt^2 + dz^2 + d\vec{x} \cdot d\vec{x},
\end{alignat}
where $\vec{x} = (x^1,\cdots,x^8)$.
The world-volume coordinates of a D2-brane are labeled by $(t,\phi,z)$.
The angle $\phi$ represents the direction along the cross section of
the supertube and $z$ lies in the transverse direction. For a supertube
with closed cross section (hereafter referred to as closed supertube),
the range of the $\phi$ is chosen as $-\pi \leq \phi \leq \pi$.
For an open supertube extending in the $x^1$-direction, we compactify it
with a radius $R_1$ and denote it as $x^1 = R_1 \phi$ and the range of
$\phi$ is taken the same as the closed one. The noncompact case is realized
by taking $R_1$ to infinity.

We consider that the position $\vec{x}$ of the D2-brane as well as the
field strengths on it depend on $t$ and $\phi$. The pullback metric and
the field strength on the D2-brane are then given by
\begin{alignat}{3}
&P[G]_{ab} =
\begin{pmatrix}
-1 + |\dot{\vec{x}}|^2 & \dot{\vec{x}} \cdot \vec{x}' & 0 \\
\dot{\vec{x}} \cdot \vec{x}' & |\vec{x}'|^2 & 0 \\
0 & 0 & 1
\end{pmatrix},
\\
&F = E dt \wedge dz + B dz \wedge d\phi,
\end{alignat}
where $\dot{x}$ and $x'$ stand for $\frac{\partial x}{\partial t}$ and
$\frac{\partial x}{\partial \phi}$.
Thus we are only considering configurations with the translational invariance
in the $z$ direction. Since turning on $E_\phi$ costs an additional
energy always, we shall not consider this possibility either.

The bosonic part of the D2-brane DBI action is evaluated as
\begin{alignat}{3}
S &= - T_2 \!\int\! dt d\phi dz \sqrt{ - \det(P[G]+\lambda\, F) }
\label{eq:actD2}
\\
&= - T_2 \!\int\! dt d\phi dz \sqrt{
(1 - |\dot{\vec{x}}|^2) (|\vec{x}'|^2 + \lambda^2 B^2) - \lambda^2 E^2
 |\vec{x}'|^2
+ (\dot{\vec{x}} \cdot \vec{x}')^2 - 2 \lambda^2 EB \dot{\vec{x}} \cdot
\vec{x}' }, \notag
\end{alignat}
where a D2-brane tension $T_2$ and $\lambda$ are written as
$T_2 = \frac{1}{(2\pi)^2\ell_s^3 g_s}$ and $\lambda = 2\pi \ell_s^2$,
respectively, in terms of the string length $\ell_s$ and coupling $g_s$.
Canonical momenta conjugate to $x^i$ and $E$ are defined as
\begin{alignat}{3}
p_i &= \frac{\delta \mathcal{L}}{\delta \dot{x}^i}
= - \frac{T_2^2}{\mathcal{L}}\big\{ \dot{x}_i (|\vec{x}'|^2 + \lambda^2 B^2)
- (\dot{\vec{x}} \cdot \vec{x}'){x_i}' + \lambda^2 EB {x_i}' \big\},
\\
\Pi &= \frac{\delta \mathcal{L}}{\delta E}
= - \frac{T_2^2}{\mathcal{L}}\big\{ \lambda^2 E |\vec{x}'|^2
+ \lambda^2 B \dot{\vec{x}} \cdot \vec{x}' \big\},
\end{alignat}
where $\mathcal{L}$ is the Lagrangian density.
As usual the Hamiltonian density is given by
\begin{alignat}{3}
\mathcal{H} &= p_i \dot{x}^i + \Pi E - \mathcal{L} \notag
\\
&= \sqrt{T_2^2|\vec{x}'|^2 + T_2^2\lambda^2 B^2 + |\vec{p}|^2
+ \frac{\Pi^2}{\lambda^2}},
\label{h1}
\end{alignat}
where use has been made of the relations
\begin{alignat}{3}
|\vec{p}|^2 &= \frac{T_2^4}{\mathcal{L}^2}\big\{ |\dot{\vec{x}}|^2
 (|\vec{x}'|^2 + \lambda^2 B^2)^2 + \lambda^4 E^2B^2 |\vec{x}'|^2 \notag
\\
&\qquad\qquad
 - (|\vec{x}'|^2 + 2\lambda^2 B^2)(\dot{\vec{x}} \cdot \vec{x}')^2
+ 2 \lambda^4 EB^3 (\dot{\vec{x}} \cdot \vec{x}') \big\} , 
\\
\Pi^2 &= \frac{T_2^4}{\mathcal{L}^2}\big\{ \lambda^4 E^2 |\vec{x}'|^4
+ \lambda^4 B^2 (\dot{\vec{x}} \cdot \vec{x}')^2
+ 2 \lambda^4 EB |\vec{x}'|^2 (\dot{\vec{x}} \cdot \vec{x}') \big\}. \notag
\end{alignat}

We are now ready to derive the BPS equations. For this purpose,
the Hamiltonian density~(\ref{h1}) is rewritten as
\begin{alignat}{3}
\mathcal{H} &= \sqrt{ \bigg( \frac{\Pi}{\lambda} + T_2 \lambda B \bigg)^2
+ \bigg( T_2|\vec{x}'| - \frac{\Pi B}{|\vec{x}'|} \bigg)^2
+ \bigg( |\vec{p}|^2 - \frac{\Pi^2 B^2}{|\vec{x}'|^2} \bigg) } .
\label{h2}
\end{alignat}
The third term in the square root is positive because of the Schwarz
inequality:
\begin{alignat}{3}
|\vec{p}|^2 - \frac{\Pi^2 B^2}{|\vec{x}'|^2} &=
\frac{T_2^4 (|\vec{x}'|^2 + \lambda^2 B^2)^2}{\mathcal{L}^2 |\vec{x}'|^2}
\big\{ |\dot{\vec{x}}|^2|\vec{x}'|^2 - (\dot{\vec{x}} \cdot \vec{x}')^2 \big\}
\geq 0.
\end{alignat}
The equality holds when $\dot{\vec{x}} \propto \vec{x}'$, which is equivalent
to $\dot{\vec{x}} = 0$ by suitable reparametrization of $\phi$.
It then follows from (\ref{h2}) that the Hamiltonian density is bounded
from below by the first term:
\begin{alignat}{3}
\mathcal{H} &\ge T \Pi + T_0 \frac{B}{2\pi},
\label{ineq1}
\end{alignat}
where $T=\frac{1}{2\pi\ell_s^2}$ is the tension of the fundamental
string and $T_0 = \frac{1}{\ell_s g_s}$ is the mass of a D0-brane.
The right hand side contains the mass density of fundamental strings
and D0-branes and precisely matches with the energy of the supertube.
The equality in (\ref{ineq1}) is saturated when both the second and third
terms in (\ref{h2}) vanish:
\begin{alignat}{3}
\Pi B = T_2 |\vec{x}'|^2 \equiv T_2 \Big(\frac{ds}{d\phi}\Big)^2, \qquad
\dot{\vec{x}} = 0,
\label{eq:BPS}
\end{alignat}
where
$ds^2 = d\vec{x} \cdot d\vec{x}$ is the line element of $\mathbb{R}^8$.
These are the BPS conditions which must be satisfied by all supertubes.
If there are flat directions at all for supertubes, they must also obey the
conditions~(\ref{eq:BPS}).

In order to examine flat directions of the supertubes keeping various
charges fixed, let us consider conserved charges of the general supertubes.
Defining
\begin{alignat}{3}
Q_1 = \frac{1}{2\pi} \int_{-\pi}^\pi d\phi\,\, \Pi, \qquad
Q_0 = \frac{1}{2\pi} \int_{-\pi}^\pi d\phi\,\, B,
\label{eq:elemag}
\end{alignat}
we find that $2\pi Q_1 \in \mathbb{Z}$ is the number of fundamental strings
dissolved in the D2-brane, and $Q_0 \int dz \in \mathbb{Z}$ is that of
D0-branes.

Other conserved charges originate from the rotational and
translational symmetries of the action (\ref{eq:actD2}).
For closed supertubes, the action possesses rotational and translational
symmetries in $\mathbb{R}^8 (\ni \!\vec{x})$.
For open supertubes in $x^1 = R_1 \phi$ direction, the action (\ref{eq:actD2})
possesses rotational and translational symmetries in $\mathbb{R}^7$.
In both cases the angular momentum $L^{ij}$ and the linear momentum $P^i$
should be conserved:
\begin{alignat}{3}
L^{ij} &= \frac{1}{2\pi} \oint_{-\pi}^\pi d\phi \,(x^i p^j - x^j p^i), \qquad
& P^i &= \frac{1}{2\pi} \oint_{-\pi}^\pi d\phi\,\, p^i,
\end{alignat}
where $i,j=1,\cdots,8$ for closed supertubes and $i,j=2,\cdots,8$ for open.
With the aid of the BPS condition (\ref{eq:BPS}),
the canonical momentum is expressed as $p^i = T_2 {x^i}'$.
The angular and linear momenta are rewritten as
\begin{alignat}{3}
L^{ij} = \frac{T_2}{\pi} \int dx^i \wedge dx^j, \qquad
P^i = \frac{T_2}{2\pi} \oint dx^i = 0. \label{eq:angmomc}
\end{alignat}
Thus, if we require the conservation of the angular momentum,
the area made by projecting the loop of the supertube onto $(i,j)$-plane
should be preserved during the deformation in the flat directions.
The linear momentum is obviously zero for any closed supertubes
since they satisfy the relation $\vec{x}(\pi) = \vec{x}(-\pi)$.

Before closing this section, we discuss the simple
example of radius changing landscape of circular supertubes with
fixed numbers of D0-branes and F-strings. Let us consider the
tubular D2-brane with uniform fluxes which is expressed as
\begin{alignat}{3}
\zeta &= x^1 + ix^2 = R e^{i\phi}, \qquad
\Pi = Q_1, \qquad B = Q_0.
\end{alignat}
Then the energy density $E$ per unit length becomes
\begin{alignat}{3}
E = \int_{-\pi}^\pi d\phi \mathcal{H} =
\sqrt{ \big( 2\pi T Q_1 + T_0 Q_0 \big)^2
+ \frac{4\pi^2}{R^2} \big( T_2 R^2 - Q_0 Q_1 \big)^2 }.
\end{alignat}
{}From this we see that the supersymmetric configuration which minimize
the energy is obtained when $Q_0 Q_1 = T_2 R^2$. Thus with fixed numbers of
D0-branes and F-strings, any area change of circular tube costs energy
for the circular supertube with uniform fluxes.

\section{Flat Directions of Circular and Planar Supertubes}

In this section, we study the flat directions in the $(x^1,x^2)$-plane
for the symmetric supertubes of circular and planar shapes.
Flat directions in the transverse directions will be discussed briefly
later on.

\subsection{Flat direction of the circular supertube}

The simplest closed configuration which satisfy the BPS equation
(\ref{eq:BPS}) with charges (\ref{eq:elemag}) is given by a supertube with
a circular cross section of radius $R$:
\begin{alignat}{3}
\zeta &= x^1 + ix^2 = R e^{i\phi}, \qquad
\Pi = Q_1, \qquad B = Q_0, \label{eq:circle}
\end{alignat}
where $-\pi \leq \phi \leq \pi$ and $Q_0 Q_1 = T_2 R^2$. Distributions of
$\Pi$ and $B$ are uniform on the world-volume of the supertube.

Let us examine flat directions of this circular supertube.
To be more explicit, we consider fluctuations around the
configuration~(\ref{eq:circle}) and impose the constraints (\ref{eq:BPS}),
(\ref{eq:elemag}) and (\ref{eq:angmomc}).
If we obtain any nontrivial solutions, these correspond to
deformations along flat directions.

Consider the fluctuations $\epsilon(\phi)$, $a(\phi)$ and $b(\phi)$ around
the circular supertube:
\begin{alignat}{3}
\zeta &= R(1+\epsilon) e^{i\phi}, \qquad
\Pi = Q_1 (1+a), \qquad B = Q_0(1+b).
\label{eq:flu}
\end{alignat}
These can be expanded as
\begin{alignat}{3}
&\epsilon = \sum_{n \in \mathbb{Z}} \epsilon_n e^{in\phi}, \qquad
a = \sum_{n \in \mathbb{Z}} a_n e^{in\phi}, \qquad
b = \sum_{n \in \mathbb{Z}} b_n e^{in\phi}.
\end{alignat}
Notice that we should impose reality conditions $\epsilon_{-n}
= \epsilon_n^\dagger, a_{-n} = a_n^\dagger, b_{-n} = b_n^\dagger$.
We require that they do not change the charges (\ref{eq:BPS}),
(\ref{eq:elemag}) and (\ref{eq:angmomc}):
\begin{alignat}{3}
& a + b + ab = 2\epsilon + \epsilon^2 + {\epsilon'}^2, \label{eq:BPSc}
\\
&a_0 = b_0 = 2\epsilon_0 + \sum_{n \in \mathbb{Z}} \epsilon_n
\epsilon_n^\dagger = 0,
\label{eq:constc}
\end{alignat}
Now we introduce a small quantity $\delta \ll 1$, and suppose that
$\epsilon, a, b \sim \mathcal{O}(\delta)$.
It follows from eq.~(\ref{eq:constc}) that the orders of magnitude of the
fluctuations are $a_n, b_n, \epsilon_n (n \geq 1) \sim \mathcal{O}(\delta)$
and $\epsilon_0 \sim \mathcal{O} (\delta^2)$.

Equations (\ref{eq:BPSc}) and (\ref{eq:constc}) are solved order
by order. We find from eq.~(\ref{eq:BPSc}) that the fluctuation $a$ is
not independent but is expressed by $\epsilon$ and $b$:
\begin{alignat}{3}
a 
= - b + 2\epsilon + b^2 - 2\epsilon b + \epsilon^2 + {\epsilon'}^2
+ \mathcal{O}(\delta^3).
\label{eq:a}
\end{alignat}
Note that $\epsilon_0$ and
$b_0$ are determined as eq.~(\ref{eq:constc}), and the remaining condition
$a_0=0$ gives a constraint on non-zero modes:
\begin{alignat}{3}
0 &= 2 \sum_{n=1} |b_n - \epsilon_n|^2
+ 2 \sum_{n=2} (n^2 - 1) |\epsilon_n|^2 + \mathcal{O}(\delta^3).
\label{cons1}
\end{alignat}
{}From eqs.~(\ref{eq:a}) and (\ref{cons1}), we learn that $\epsilon_1$ is
the only independent fluctuation and
\begin{alignat}{3}
\epsilon &= \epsilon_1 e^{i\phi} + \epsilon_1^\dagger e^{-i\phi}
+ \mathcal{O}(\delta^2), \notag
\\
a &= \epsilon_1 e^{i\phi} + \epsilon_1^\dagger e^{-i\phi} +
\mathcal{O}(\delta^2),
\\
b &= \epsilon_1 e^{i\phi} + \epsilon_1^\dagger e^{-i\phi} + \mathcal{O}(
\delta^2).  \notag
\end{alignat}

Repeating the same procedure up to the order of $\mathcal{O}(\delta^5)$,
we find that the fluctuations are expressed as
\begin{alignat}{3}
\epsilon &= \epsilon_1 e^{i\phi} + \epsilon_1^\dagger e^{-i\phi}
- |\epsilon_1|^2 + \tfrac{1}{2} \epsilon_1^2 e^{2i\phi} + \tfrac{1}{2}
\epsilon_1^{\dagger 2} e^{-2i\phi} + \mathcal{O}(\delta^3), \notag \\
a &= \epsilon_1 e^{i\phi} + \epsilon_1^\dagger e^{-i\phi}
+ \mathcal{O}(\delta^3),
\label{eq:flu2}
\\
b &= \epsilon_1 e^{i\phi} + \epsilon_1^\dagger e^{-i\phi}
+ \mathcal{O}(\delta^3). \notag
\end{alignat}
This procedure can be repeated to determine $\epsilon$, $a$
and $b$ to any order we want.

We are now going to show that the fluctuations (\ref{eq:flu2})
represent the trivial flat direction corresponding to the translation
of the circular supertube without deformation in $(x^1,x^2)$-plane.
Consider a circle whose center is located at $2R\epsilon_1^\dagger$ in
$(x^1,x^2)$-plane:
\begin{alignat}{3}
R^2 &= |\zeta - 2R\epsilon_1^\dagger|^2, \qquad \zeta = R(1+\epsilon)e^{i\phi}.
\end{alignat}
This can be solved for real $\epsilon$ as
\begin{alignat}{3}
\epsilon = -1 + \epsilon_1 e^{i\phi} + \epsilon_1^\dagger e^{-i\phi}
+ \sqrt{1 - 2|\epsilon_1|^2 + \epsilon_1^2 e^{2i\phi}
+ \epsilon_1^{\dagger 2} e^{-2i\phi}}.
\end{alignat}
When expanded in $\epsilon_1$, this precisely reproduces (\ref{eq:flu2}).
Also $\Pi$ and $B$ are proportional to the line element
$|\frac{d\zeta}{d\phi}|$, so the fluctuations $a$ and $b$ are given by
\begin{alignat}{3}
a = b = \frac{1}{R} \Big|\frac{d\zeta}{d\phi} \Big| - 1
= \frac{\epsilon_1 e^{i\phi} + \epsilon_1^\dagger e^{-i\phi}}
{\sqrt{1 - 2|\epsilon_1|^2 + \epsilon_1^2 e^{2i\phi} + \epsilon_1^{\dagger 2}
e^{-2i\phi}}}.
\end{alignat}
When expanded in $\epsilon_1$, these again reproduce (\ref{eq:flu2}).
As mentioned in the previous section, the action (\ref{eq:actD2})
possesses the translational symmetry in $\mathbb{R}^8$. Our analysis indicates
that the only allowed flat directions are these trivial translations and there
is no other flat direction.

\subsection{Flat direction of the planar supertube}

We next investigate the flat directions of the planar supertube lying in
$x^1 = R_1 \phi$. The uncompactified case is realized by sending $R_1$ to
$\infty$. The planar supertube which satisfies the BPS eq. (\ref{eq:BPS})
with charges (\ref{eq:elemag}) is expressed as
\begin{alignat}{3}
x^2 = 0, \qquad \Pi = Q_1, \qquad B = Q_0,
\end{alignat}
where $Q_0 Q_1 = T_2 R_1^2$. We consider fluctuations around this configuration
\begin{alignat}{3}
x^2 &= R_1 \epsilon, \qquad \Pi = Q_1 (1+a), \qquad B = Q_0(1+b),
\end{alignat}
which are expanded as
\begin{alignat}{3}
&\epsilon = \sum_{n=-\infty}^\infty \epsilon_n e^{in\phi}, \qquad
a = \sum_{n=-\infty}^\infty a_n e^{in\phi}, \qquad
b = \sum_{n=-\infty}^\infty b_n e^{in\phi},
\end{alignat}
where $a_{-n} = a_n^\dagger$, mutatis mutandis. Taking into account
the BPS eq.~(\ref{eq:BPS}) and the conserved charges~(\ref{eq:elemag}),
we obtain the conditions for the fluctuations
\begin{alignat}{3}
& a + b + ab = {\epsilon'}^2,
\label{eq:BPSp} \\
&a_0 = b_0 = 0.
\label{eq:constp}
\end{alignat}
The conservation laws (\ref{eq:angmomc}) are trivially satisfied.

Now we show that there are no flat directions except for the trivial
translation. Let us introduce a small quantity $\delta \ll 1$, and assume
that  $\epsilon, a, b \sim \mathcal{O}(\delta)$. Equations~(\ref{eq:BPSp})
and (\ref{eq:constp}) are then solved order by order as before.
It follows from eq. (\ref{eq:BPSp}) that the fluctuation $a$ is represented
by $\epsilon$ and $b$ as
\begin{alignat}{3}
a 
= - b + {\epsilon'}^2 + b^2 + \mathcal{O}(\delta^3),
\label{eq:ap}
\end{alignat}
We know from eq.~(\ref{eq:constp})
that $b_0 = 0$ and the remaining constraint $a_0 = 0$ is expressed as
\begin{alignat}{3}
0 &= 2 \sum_{n=1} |b_n|^2
+ 2 \sum_{n=1} n^2|\epsilon_n|^2 + \mathcal{O}(\delta^3).
\label{cons2}
\end{alignat}
Equations~(\ref{eq:ap}) and (\ref{cons2}) lead to
\begin{alignat}{3}
\epsilon = \epsilon_0 + \mathcal{O}(\delta^2), \qquad
a = \mathcal{O}(\delta^2), \qquad
b = \mathcal{O}(\delta^2).
\end{alignat}
Repeating the above procedure to the higher order, we find that the
allowed fluctuations are given exactly by
\begin{alignat}{3}
\epsilon = \epsilon_0, \qquad a = 0, \qquad
b = 0.
\end{alignat}
This is nothing but the trivial translation of the planar supertube
in the $x^2$-direction. Thus we conclude that there are no nontrivial flat
directions for planar supertubes.

\section{Flat Directions of Supertubes of General Shape}

In this section we study flat directions of supertubes of general shape
with fixed fundamental string charge $Q_1$, D0-brane charge $Q_0$ and angular
momentum $J$.

We first discuss closed supertubes.
The perimeter of the cross section is denoted by $2\pi L$,
and $J$ and $L$ are written as
\begin{alignat}{3}
J = \frac{T_2}{\pi} \int dx^1 \wedge dx^2, \qquad
L = \frac{1}{2\pi} \int_{-\pi}^\pi d\phi \frac{ds}{d\phi},
\end{alignat}
where $ds$ is the line element of the cross section.

Let us first consider the allowed range of $L$ for given $Q_0, Q_1$ and $J$.
The lower bound for $L$ is obtained by noting the corollary of the isoperimetric theorem.
It is known that for the fixed area $\frac{\pi}{T_2}J$, the circle gives
the shortest perimeter. Therefore we obtain
\begin{alignat}{3}
\sqrt{\frac{J}{T_2}} \leq L.
\end{alignat}
The equality holds when the cross section becomes circular.

Using the Schwarz inequality, the upper bound on $L$ is found to be
\begin{alignat}{3}
L =\frac{1}{2\pi}\int_{-\pi}^\pi d\phi \sqrt{|\vec x'|^2}
&\leq \sqrt{\bigg( \frac{1}{2\pi} \int_{-\pi}^\pi d\phi \frac{T_2^{2/3}
|\vec{x}'|^2}{B} \bigg) \bigg( \frac{1}{2\pi} \int_{-\pi}^\pi d\phi
\frac{B}{T_2^{2/3}} \bigg)} \notag \\
&= \sqrt{\bigg( \frac{1}{2\pi} \int_{-\pi}^\pi d\phi \frac{\Pi}{T_2^{1/3}}
\bigg) \bigg( \frac{1}{2\pi} \int_{-\pi}^\pi d\phi \frac{B}{T_2^{2/3}} \bigg)}
\notag \\
&= \sqrt{\frac{Q_0 Q_1}{T_2}},
\label{sch}
\end{alignat}
where use has been made of eq.~(\ref{eq:BPS}) in going to the second line,
and (\ref{eq:elemag}) in the third line.
The equality holds if
\begin{alignat}{3}
\Pi = \sqrt{\frac{Q_1 T_2}{Q_0}} \frac{ds}{d\phi}, \qquad
B = \sqrt{\frac{Q_0 T_2}{Q_1}} \frac{ds}{d\phi}, \label{eq:base}
\end{alignat}
which means that $\Pi$ and $B$ are distributed uniformly along the cross
section. We thus find that the range of $L$ is bounded as
\begin{alignat}{3}
\sqrt{\frac{J}{T_2}} \leq L \leq \sqrt{\frac{Q_0 Q_1}{T_2}}.
\label{range}
\end{alignat}
When $J = Q_0 Q_1$, the supertube is uniquely circular with uniform electric
and magnetic fluxes.
When $J > Q_0 Q_1$, we have no supertube solutions.

For $J < Q_0 Q_1$, let us show that there are infinite flat directions in
the whole range~(\ref{range}). For this purpose, we choose the base
configuration, labeled by $t=0$, of the supertube with fixed $Q_0, Q_1, J$
and maximum length:
\begin{alignat}{3}
L_0 = \sqrt{\frac{Q_0 Q_1}{T_2}}, \qquad
\Pi_0 = \sqrt{\frac{Q_1 T_2}{Q_0}} \frac{ds_0}{d\phi}, \qquad
B_0 = \sqrt{\frac{Q_0 T_2}{Q_1}} \frac{ds_0}{d\phi}.
\end{alignat}
The shape of the cross section can be chosen as the ellipse for example,
and the area is equal to $\frac{\pi J}{T_2}$.
The other end configuration, labeled by $t=1$, is taken to be a supertube
with fixed $Q_0, Q_1$ and $J$ characterized by
\begin{alignat}{3}
L_1 = \frac{1}{2\pi} \int_{-\pi}^\pi d\phi \frac{ds_1}{d\phi}, \qquad
\Pi_1 = \sqrt{\frac{Q_1 T_2}{Q_0}} \frac{ds_1}{d\phi} f_1, \qquad
B_1 = \sqrt{\frac{Q_0 T_2}{Q_1}} \frac{ds_1}{d\phi} f_1^{-1},
\label{f0}
\end{alignat}
whose shape of the cross section can be chosen arbitrary with length
$2\pi L_1$, but its enclosed area should be kept the same
$\frac{\pi J}{T_2}$. From the requirement that the charges~(\ref{eq:elemag})
be conserved, the function $f_1$ must satisfy the conditions
\begin{alignat}{3}
2\pi L_0 = \int_0^{2\pi L_1} ds_1 f_1(s_1) =
\int_0^{2\pi L_1} ds_1 f_1^{-1}(s_1).
\label{f1}
\end{alignat}

To find such $f_1$ with full generality, let us consider the function
$h_1(s_1)$ which satisfies
\begin{alignat}{3}
2\pi L_0 = \int_0^{2\pi L_1} ds_1 \sqrt{1 + \Big(\frac{dh_1}{ds_1}\Big)^2},
\label{curve}
\end{alignat}
with $h_1(2\pi L_1) = h_1(0) = 0$. Geometrically this means that we
have a curve $(s_1,h_1(s_1))$ whose length for $0 \leq s_1 \leq 2\pi L_1$
is equal to $2\pi L_0 (\geq 2\pi L_1)$ with both ends fixed.
We should note that this is a condition that defines $h_1$ and
the actual length of the deformed curve is $2\pi L_1$.
We find that the generic $f_1$ satisfying (\ref{f1}) is given by
$\frac{d h_1}{d s_1} = \frac{1}{2}(f_1 - f_1^{-1})$ or
\begin{alignat}{3}
f_1=\sqrt{1+ \Big(\frac{dh_1}{ds_1}\Big)^2} + \frac{dh_1}{ds_1}.
\end{alignat}
This prescribes how the fluxes are distributed keeping the BPS
conditions~(\ref{eq:BPS}).

Now we consider a cross section, labeled by $t$, which continuously deforms
between the base configuration and the other end configuration.
Its perimeter and the area are $2\pi L_t$ and $\frac{\pi J}{T_2}$,
respectively. The line element for $L_t$ is denoted as $ds_t$.
We choose $\Pi_t$ and $B_t$ as
\begin{alignat}{3}
\Pi_t (s_t) &= \sqrt{\frac{Q_1 T_2}{Q_0}} \frac{ds_t}{d\phi}
\bigg( \frac{dh_t}{ds_t} + \sqrt{1 + \Big(\frac{dh_t}{ds_t}\Big)^2} \bigg),
\notag  \\
B_t (s_t) &= \sqrt{\frac{Q_0 T_2}{Q_1}} \frac{ds_t}{d\phi}
\bigg( - \frac{dh_t}{ds_t} + \sqrt{1 + \Big(\frac{dh_t}{ds_t}\Big)^2} \bigg),
\end{alignat}
where the function $h_t(s_t)$ is arranged to satisfy
\begin{alignat}{3}
2\pi L_0 = \int_0^{2\pi L_t} ds_t \sqrt{1 + \Big(\frac{dh_t}{ds_t}\Big)^2},
\qquad
h_t(2\pi L_t) = h_t(0) = 0. \label{eq:h_t}
\end{alignat}
It is then easy to see that $\Pi_t$ and $B_t$ obey the conservation law.
Thus we can always find $h_t$ which continuously deforms between
$h_0=0$ and $h_1$. This establishes our desired result.

Flat directions for general open supertube can be treated similarly.
Suppose that the $x^1$-direction is compactified and consider the open
supertube which extends in this direction $x^1 = R_1 \phi$.
Then the length of the cross section $L$ is bounded as
\begin{alignat}{3}
R_1 \leq L \leq \sqrt{\frac{Q_0 Q_1}{T_2}}.
\label{eq:range2}
\end{alignat}
When $T_2 R_1^2 = Q_0 Q_1$, the supertube is uniquely planar with uniform
electric and magnetic fluxes.
When $T_2 R_1^2 > Q_0 Q_1$, we have no supertube solutions.
For $T_2 R_1^2 < Q_0 Q_1$, there are infinite flat directions in
the whole range~(\ref{eq:range2}).

Finally let us briefly discuss the case where the cross section of the
supertube makes use of remaining transverse directions too. While
preserving the angular momentums $L_{ij}$, the cross sectional contour
may vary including the deformations in the transverse directions.
The total length of the curve should be limited by the inequality
(\ref{sch}). Then the distribution of $B$ and $\Pi$ along the
curve may be given by (\ref{f0}) and (\ref{f1})
by the same way as before. It is clear that these are
moduli deformations in the phase space with the transverse
dimensions included.

\section{Conclusions}

In this note, we have investigated the phase space landscape of
the supertubes. The arbitrariness of the cross sectional shape and
the density of D0 branes are allowed within the region preserving 1/4
of supersymmetries of IIA string theory. We have shown that, with
the fixed numbers of D0 and F1, increasing or reducing the cross
sectional area makes the energy larger always. Thus these
deformations are not moduli. We have also shown that there are no
flat directions for the circular supertubes with uniform D0 density
and angular momentum except the trivial overall translation
in the transverse space. This is the case where $Q_0 Q_1=J$.
When $Q_0 Q_1 > J$, there are many flat directions of supertubes
with fixed angular momentum. We figure out explicitly all the phase
space flat directions, which form the   phase moduli space.

Having the statistical applications in mind, we are ultimately interested
in the phase space volume of the 1/4 BPS supertubes with all the
conserved quantities fixed. Though we identified all the detailed
structures of the   phase moduli space, we are not able to obtain the
expression for the volume. Of course the logarithm of this volume
corresponds to the zero temperature entropy of the system with fixed
charges. Identification of the volume with quadratic approximation
around circular supertubes is done in Ref.~\cite{marolf}.
The computation of the volume without approximation requires further studies.

The above counting may be compared to the horizon area of the extremal
supertube solutions~\cite{emparan,emparan2,bena2,bena3,lunin}. Furthermore
near BPS counting of the volume will correspond to the
horizon area of near extremal black supertube~\cite{bena2}.
Details of the correspondence will be of much interest.

\vspace{.7cm}
\noindent{\large\bf Acknowledgment}

We are grateful to Kyung Kyu Kim for the contributions at the
earlier stage of this work
and Seok Kim for useful discussions and conversations.
DB is supported in part by KOSEF
ABRL R14-2003-012-01002-0 and KOSEF
R01-2003-000-10319-0.
The work of YH was supported in part by
a Grant-in-Aid for JSPS fellows, and
NO was supported in part
by a Grant-in-Aid for Scientific Research Nos. 16540250.

\end{document}